\documentclass[fleqn,twoside]{article}
\usepackage{espcrc2}

\usepackage{graphicx}
\usepackage[figuresright]{rotating}

\newcommand{\AmS}{{\protect\the\textfont2
  A\kern-.1667em\lower.5ex\hbox{M}\kern-.125emS}}

\title{Leptogenesis and CP Violation in Neutrino Oscillations}

\author{Zhi-zhong Xing \address{Institute of High Energy Physics,
	P.O. Box 918 (4), Beijing 100039, China}}%
       
\begin{document}

\begin{abstract}
Assuming the seesaw and leptogenesis mechanisms,
I give some general remarks on possible connection or disconnection 
between CP violation in heavy Majorana neutrino decays
and that in neutrino oscillations. A simple but predictive ansatz 
is proposed to simultaneously interpret the observed baryon 
asymmetry of the universe and oscillations of solar and atmospheric 
neutrinos.
\end{abstract}

\maketitle

The density of baryons compared to that of photons in the universe
is extremely small: $\eta \equiv n^{~}_{\rm B}/n^{~}_\gamma =
(2.6 - 6.3) \times 10^{-10}$, extracted from the Big-Bang
nucleosynthesis \cite{PDG}. This tiny quantity is related to
the observed matter-antimatter or baryon-antibaryon asymmetry of 
the universe,
\begin{equation}
Y_{\rm B} \equiv \frac{n^{~}_{\rm B}}{\bf s} \approx 
\frac{\eta}{7.04} = (3.7 - 8.9) \times 10^{-11} ,
\end{equation}
where $\bf s$ denotes the entropy density. In order to produce a net baryon 
asymmetry in the standard Big-Bang model, three Sakharov necessary 
conditions must be satisfied \cite{Sakharov}: 
(a) baryon number nonconservation, (b) C and CP violation, and 
(c) a departure from thermal equilibrium. Among several  
interesting and viable baryogenesis scenarios proposed in the literature,
Fukugita and Yanagida's leptogenesis mechanism \cite{FY} has 
attracted a lot of attention -- due partly to the fact that neutrino
physics is entering a flourishing era.

Indeed the Super-K \cite{SK} and SNO \cite{SNO} data have 
provided very convincing evidence that neutrinos are massive and 
lepton flavors are mixed -- a kind of new physics which can only be
understood beyond the standard electroweak model.
A simple extension of the standard model is to include one
right-handed neutrino in each of three lepton families, while 
the Lagrangian of electroweak interactions keeps invariant under 
the $\rm SU(2)_{\rm L} \times U(1)_{\rm Y}$ gauge 
transformation. In this case, the Yukawa interactions are described by
\begin{equation}
- {\cal L}_{\rm Y} = \overline{l}_{\rm L} \tilde{\phi} Y_l e^{~}_{\rm R}
+ \overline{l}_{\rm L} \phi Y_\nu \nu^{~}_{\rm R} +
\frac{1}{2} \overline{\nu^{\rm c}_{\rm R}} M_{\rm R} \nu^{~}_{\rm R} 
+ {\rm h.c} , ~
\end{equation}
where $l_{\rm L}$ denotes the left-handed lepton doublet, $e^{~}_{\rm R}$
and $\nu^{~}_{\rm R}$ stand respectively for the right-handed charged lepton
and Majorana neutrino singlets, and $\phi$ is the standard-model Higgs 
doublet. The lepton number violation induced by the third term of 
${\cal L}_{\rm Y}$ allows decays of the 
heavy right-handed Majorana neutrinos $N_i$ (for $i=1,2,3$) to occur:
\begin{equation}
N_i \rightarrow l + \phi^\dagger ~~~ {\rm vs} ~~~
N_i \rightarrow l^{\rm c} + \phi \; .
\end{equation}
As each decay can happen both at the tree level and at the
one-loop level (via the self-energy and vertex corrections), the
interference between tree-level and one-loop decay amplitudes
may give rise to a CP-violating asymmetry $\varepsilon_i$ between the two
CP-conjugated processes in Eq. (3) \cite{FY}. 
If the masses of three heavy Majorana 
neutrinos $N_i$ are hierarchical ($M_1 \ll M_2 \ll M_3$) and the 
interactions of $N_1$ are in thermal equilibrium when $N_2$ and $N_3$ decay, 
the asymmetries produced by $N_2$ and $N_3$ can be erased before 
$N_1$ decays. The CP asymmetry $\varepsilon_1$ produced by
the out-of-equilibrium decay of $N_1$ survives, and it results in a
lepton-antilepton asymmetry $Y_{\rm L} \equiv n^{~}_{\rm L}/{\bf s}
= \varepsilon_1 d/g^{~}_*$, where $g^{~}_* \sim 100$ is an effective number
characterizing the relativistic degrees of freedom which contribute to 
the entropy {\bf s}, and $d$ accounts for the dilution effects induced by
the lepton-number-violating wash-out processes. Finally the lepton
asymmetry $Y_{\rm L}$ is converted into a net baryon asymmetry
$Y_{\rm B}$ through the $({\rm B + L})$-violating sphaleron 
processes \cite{Kuzmin}: 
\begin{equation}
Y_{\rm B} = \frac{c}{c - 1} Y_{\rm L} = \frac{c}{c - 1} \cdot
\frac{d}{g^{~}_*} \varepsilon_1 \; ,
\end{equation}
where $c=(8{\rm N}_f + 4{\rm N}_\phi)/(22{\rm N}_f + 13{\rm N}_\phi)$
with ${\rm N}_f$ being the number of fermion families and 
${\rm N}_\phi$ being the number of Higgs doublets. Taking
${\rm N}_f = 3$ and ${\rm N}_\phi =1$ for example, we obtain
$c \approx 1/3$. 

After spontaneous symmetry breaking, we get the charged lepton
mass matrix $M_l = Y_l \langle \phi \rangle$ and the Dirac neutrino
mass matrix $M_{\rm D} = Y_\nu \langle \phi \rangle$ from ${\cal L}_{\rm Y}$,
in addition to the right-handed Majorana neutrino mass matrix $M_{\rm R}$.
The scale of $M_l$ and $M_{\rm D}$ is characterized 
by the gauge symmetry breaking scale 
$v \equiv \langle \phi \rangle \approx 175$ GeV, but that of 
$M_{\rm R}$ may be much higher than
$v$, because right-handed neutrinos are $\rm SU(2)_{\rm L}$ singlets
and their mass term is not subject to the electroweak symmetry
breaking. As a consequence, the light (and essentially left-handed) 
neutrino mass matrix $M_\nu$ can be obtained via the seesaw 
mechanism \cite{Seesaw}:
\begin{equation}
M_\nu \approx - M_{\rm D} M^{-1}_{\rm R} M^{\rm T}_{\rm D} \; .
\end{equation}
Note that lepton flavor mixing at low energy scales 
stems from a nontrivial mismatch between the diagonalizations of $M_\nu$ and 
$M_l$, while the baryon asymmetry at high energy scales depends on complex 
$M_{\rm D}$ and $M_{\rm R}$ in the leptogenesis scenario. To see these
points more clearly, let us diagonalize three of the four lepton mass matrices:
$U^\dagger_l M_l \tilde{U}_l = {\rm Diag} \{ m_e, m_\mu, m_\tau \}$ and
$U^\dagger_\nu M_\nu \tilde{U}^*_\nu = {\rm Diag} \{ m_1, m_2, m_3 \}$
as well as $U^\dagger_{\rm R} M_{\rm R} \tilde{U}^*_{\rm R} 
= {\rm Diag} \{ M_1, M_2, M_3 \}$.
Then the lepton flavor mixing matrix at low energy scales
is given by $V = U^\dagger_l U_\nu$, and the leptonic CP violation
in neutrino oscillations is measured by the Jarlskog parameter 
$\cal J$ defined through 
\begin{equation}
{\rm Im} \left (V_{\alpha i} V_{\beta j} V^*_{\alpha j} V^*_{\beta i} \right )
= {\cal J} \sum_{\gamma, k} \left (\epsilon^{~}_{\alpha\beta\gamma} \cdot
\epsilon^{~}_{ijk} \right ) ,
\end{equation}
where $(\alpha, \beta)$ run over $(e, \mu, \tau)$ and $(i, j)$ 
run over $(1,2,3)$. On the other hand, the CP asymmetry $\varepsilon_1$ 
between $N_1 \rightarrow l + \phi^\dagger$ and 
$N_1 \rightarrow l^{\rm c} + \phi$ decays at high energy scales 
can be given as \cite{Xing02}
\begin{eqnarray}
\varepsilon_1 = - \frac{3}{16\pi v^2} \cdot
\frac{M_1}{\left [U^{\rm T}_{\rm R} M^\dagger_{\rm D} M_{\rm D} 
U^*_{\rm R} \right ]_{11}} 
\nonumber \\
\cdot \sum^3_{j=2} \frac{{\rm Im} 
\left ( \left [U^{\rm T}_{\rm R} M^\dagger_{\rm D}
M_{\rm D} U^*_{\rm R} \right ]_{1j} \right )^2}{M_j} \; ,
\end{eqnarray}
where the mass hierarchy of three heavy Majorana neutrinos 
($M_1 \ll M_2 \ll M_3$) has been assumed. We see that $\cal J$ depends on
$U_l$ and $U_\nu$ or equivalently on $M_l$ and $M_\nu$,
while $\varepsilon_1$ depends on $M_{\rm D}$ and $M_{\rm R}$
(or $U_{\rm R}$ and $M_i$). The only possible relationship between
$\cal J$ and $\varepsilon_1$ is due to the seesaw mechanism in Eq. (5),
which links $M_\nu$ to $M_{\rm D}$ and $M_{\rm R}$. Therefore one can
conclude that there is no direct connection between CP violation in
heavy Majorana neutrino decays ($\varepsilon_1$) and that in
neutrino oscillations ($\cal J$). Such a general conclusion was
first drawn by Buchm$\rm\ddot{u}$ller and Pl$\rm\ddot{u}$macher \cite{BP}.
Recently a few other authors have carried out some
more delicate analyses and reached the same conclusion \cite{Others}. 

Depending on the specific flavor basis that we choose in model building,
$\cal J$ and $\varepsilon_1$ can either be
completely disconnected or maximally connected. To illustrate,
let us consider two extreme cases:

(1) In the basis where $U_\nu = {\bf 1}$ holds (i.e., 
$V = U^\dagger_l$ -- lepton flavor mixing and CP violation in neutrino
oscillations arise solely from the charged lepton sector \cite{FX96}), 
we find that $\varepsilon_1$ has nothing to do with $V$ or $\cal J$. 
In this special case, less fine-tuning is expected in building a
phenomenological model which can simultaneously interpret the
baryon asymmetry of the universe and lepton flavor mixing at low
energy scales.

(2) In the basis where both $U_{\rm R} = {\bf 1}$ and 
$U_l = {\bf 1}$ hold (i.e.,
$V = U_\nu$ -- lepton flavor mixing and CP violation in neutrino
oscillations arise solely from the neutrino sector), we find that
$\varepsilon_1$ can indirectly be linked to $V$ and $\cal J$
through the seesaw relation in Eq. (5).
It is worth remarking that the correlation between high- and low-energy
observable quantities requires quite nontrivial attempts in
model building. In particular, the textures of $M_{\rm R}$ and $M_{\rm D}$ 
have to be carefully chosen or fine-tuned to guarantee acceptable
agreement between the model predictions and the observational or
experimental data. 

\vspace{0.35cm}

In the second part of this talk, I propose a phenomenological ansatz to 
simultaneously interpret the observed baryon asymmetry of the universe and 
neutrino oscillations. Our simple ansatz \cite{Xing02} is essentially a 
non-SO(10) modification of the Buchm$\rm\ddot{u}$ller-Wyler ansatz \cite{BW},
but it has more powerful predictability and its predictions are
in better agreement with current data.

First of all, we assume $M_l$ and $M_{\rm D}$ 
to be symmetric matrices, just like $M_{\rm R}$. Second, we
assume that the (1,1), (1,3) and (3,1) elements of $M_l$,
$M_{\rm D}$ and $M_{\rm R}$ are all vanishing in a specific flavor basis, 
like a phenomenologically-favored 
texture of quark mass matrices $M_{\rm u}$ and $M_{\rm d}$ \cite{FX95}.
Third, we assume that the non-zero elements of $M_{\rm D}$ and $M_l$ 
can be expanded in terms of the Wolfenstein parameter 
$\lambda \approx 0.22$. To be explicit, we 
conjecture that $M_{\rm D}$ and $M_l$ have the following patterns:
\begin{eqnarray}
\frac{M_{\rm D}}{m^{~}_0} = \left [ \matrix{
0	& \hat{\lambda}^3	& 0 \cr
\hat{\lambda}^3	& x \hat{\lambda}^2	& \hat{\lambda}^2 \cr
0	& \hat{\lambda}^2 	& e^{i\zeta} \cr} \right ]  , 
\frac{M_l}{m^{~}_\tau} = \left [ \matrix{
0	& \lambda^4	& 0 \cr
\lambda^4	& y \lambda^2	& \lambda^3 \cr
0	& \lambda^3 	& 1 \cr} \right ]
\nonumber 
\end{eqnarray}
with $m_0 \approx v$, $\hat{\lambda} \equiv \lambda e^{i\omega}$ and
$(x, y)$ being real and positive coefficients of ${\cal O}(1)$. It is
easy to check that three mass eigenvalues of $M_{\rm D}$ have the hierarchy
$\lambda^4 : \lambda^2 : 1$, and those of $M_l$ have the hierarchy 
compatible with our experimental data on $m_e$, $m_\mu$ and $m_\tau$.
As the (1,1), (1,3) and (3,1) elements of both $M_{\rm R}$ and 
$M_{\rm D}$ are vanishing, $M_\nu$ must have the 
same texture zeros via the seesaw relation in Eq. (5) \cite{FX99}.
To generate a sufficiently large mixing angle in the 
$\nu_\mu$-$\nu_\tau$ sector, (2,3), (3,2) and (3,3) elements of 
$M_\nu$ should be comparable in magnitude.
This requirement is actually strong enough to constrain the texture 
of $M_{\rm R}$ in a quite unique way \cite{BW}. For our purpose, we obtain
\begin{eqnarray}
\frac{M_{\rm R}}{M_0} = \left [ \matrix{
0	& \lambda^5	& 0 \cr
\lambda^5	& z \lambda^4	& \lambda^4 \cr
0	& \lambda^4	& 1 \cr} \right ] , 
\frac{M_\nu}{m'_0} = \left [ \matrix{
0	& \hat{\lambda}		& 0 \cr
\hat{\lambda}	& z'	& 1 \cr
0	& 1	& e^{i2\varphi} \cr} \right ] ,
\nonumber
\end{eqnarray}
where $M_0 \gg v$, $m'_0 = v^2/M_0$, 
$z$ is a real and positive coefficient of ${\cal O}(1)$, 
$z' \equiv 2x - ze^{i\omega}$ with $|z'| \sim {\cal O}(1)$,
and $2\varphi \equiv 2\zeta - 5\omega$. Note that
an overall phase factor $e^{i(5\omega-\pi)}$ has
been omitted from the right-hand side of $M_\nu$, 
since it has no contribution to lepton flavor mixing
and CP violation. To generate a large mixing angle
in the $\nu_e$-$\nu_\mu$ sector, the condition 
$| z' e^{i2\varphi} - 1 | \equiv \delta \sim {\cal O}(\lambda)$
must be satisfied. There exists an
interesting parameter space, in which \cite{Xing02}
$x = 1/\sqrt{2}$, $z = 1 + \sqrt{2} \lambda$ and
$\zeta = - \omega = \pi/4$. One may check that 
$\delta = \sqrt{2} \lambda$ holds in this parameter space. 

As mentioned above, complex $M_\nu$ can be diagonalized by 
a unitary matrix $U_\nu$, On the other hand, the strong hierarchy of 
real $M_l$ under consideration implies that its contribution to lepton 
flavor mixing is small and negligible. Then we arrive at $V \approx U_\nu$, 
which links the neutrino mass eigenstates ($\nu_1, \nu_2, \nu_3$) 
to the neutrino flavor eigenstates ($\nu_e, \nu_\mu, \nu_\tau$).
Current data on solar and atmospheric
neutrino oscillations suggest $|V_{e3}| \ll 1$, 
$|V_{e1}| \sim |V_{e2}|$ and $|V_{\mu 3}| \sim |V_{\tau 3}|$. 
Hence a parametrization of $V$ needs two big
mixing angles ($\theta_x$ and $\theta_y$) and one small mixing angle 
($\theta_z$), in addition to a few complex phases. Following 
Ref. \cite{Xing02}, we obtain 
\begin{eqnarray}
\tan 2\theta_x \approx 2\sqrt{2} \frac{\lambda}{\delta} \; , ~
\tan 2\theta_y \approx \frac{2}{\delta} \; , ~
\tan 2\theta_z \approx \frac{\lambda}{\sqrt{2}} \; . 
\nonumber
\end{eqnarray}
Taking $\delta = \sqrt{2}\lambda$, we explicitly obtain
$\theta_x \approx 31.7^\circ$, 
$\theta_y \approx 40.6^\circ$ and
$\theta_z \approx 4.4^\circ$, favored by current experimental data.
In addition, the masses of three light neutrinos are given by
\begin{eqnarray}
\frac{m_1}{m_3} \approx \frac{\lambda \tan\theta_x}{2\sqrt{2}} \; , ~~ 
\frac{m_2}{m_3} \approx \frac{\lambda \cot\theta_x}{2\sqrt{2}} \; , ~~
m_3 \approx \frac{2 m^2_0}{M_0} \; .
\nonumber
\end{eqnarray}
We see that a normal neutrino mass hierarchy 
$m_1 : m_2 : m_3 \sim \lambda : \lambda : 1$ shows up.
Then the absolute value of $m_3$ can be determined from the observed 
mass-squared difference of atmospheric neutrino oscillations 
$\Delta m^2_{\rm atm} \equiv |m^2_3 - m^2_2| \approx m^2_3$.
Using $\Delta m^2_{\rm atm} = (1.6 - 3.9) \times 10^{-3} ~ {\rm eV}^2$
\cite{Shiozawa}, we obtain
$m_3 \approx \sqrt{\Delta m^2_{\rm atm}} \approx 
(4.0 - 6.2) \times 10^{-2} ~ {\rm eV}$.
Given $m_0 \approx v$ for the Dirac neutrino mass matrix $M_{\rm D}$, 
the mass scale of three heavy Majorana neutrinos turns out 
to be
$M_0 \approx 2 v^2/m_3 \approx (4.9 - 7.6) \times 10^{14} ~ {\rm GeV}$,
which is quite close to the scale of grand unified
theories $\Lambda_{\rm GUT} \sim 10^{16}$ GeV.

Note that our ansatz predicts a very small value for the effective mass term 
of the neutrinoless double-$\beta$ decay:
$\langle m\rangle_{ee} \approx \lambda^2 m_3/8 \approx (2.4 - 3.8) 
\times 10^{-4} ~ {\rm eV}$, which is much lower than the present 
experimental upper bound ($\langle m\rangle_{ee} < 0.35$ eV at the 
$90\%$ C.L.\cite{Beta}) and seems hopeless to be detected 
in practice. We also obtain the Jarlskog parameter, which measures
CP and T violation in neutrino oscillations,
as ${\cal J} \approx \lambda /(4\sqrt{10}) \approx 2\%$. 
Leptonic CP violation at the percent level could be measured in the 
future at neutrino factories.

The symmetric mass matrix $M_{\rm R}$ 
can be diagonalized by a unitary matrix $U_{\rm R}$,
as pointed out above. To leading order, we find
$M_1 \approx \lambda^6 M_0/z$, $M_2 \approx z \lambda^4 M_0$ and
$M_3 \approx M_0$ as well as 
$U_{\rm R 11} \approx i$, $U_{\rm R 22} \approx U_{\rm R 33} \approx 1$,
$U_{\rm R 12} \approx i U_{\rm R 21} \approx \lambda/z$,
$U_{\rm R 13} \approx 0$, $U_{\rm R 31} \approx i\lambda^5/z$, and
$U_{\rm R 23} \approx -U_{\rm R 32} \approx \lambda^4$.
One can see that the masses of three heavy Majorana neutrinos 
perform a strong hierarchy. Note that
$\{ M_1, M_2, M_3 \} \approx \{5.2 \times 10^{10}, 
1.8 \times 10^{12}, 6.0 \times 10^{14} \} ~ {\rm GeV}$,
if we typically take $M_0 = 6.0 \times 10^{14} ~ {\rm GeV}$ and 
$z = 1 + \sqrt{2}\lambda$.

A CP-violating asymmetry ($\varepsilon_1$) may result from the interference 
between tree-level and one-loop amplitudes of the decay of the {\it lightest} 
heavy Majorana neutrino $N_1$, as already presented in Eq. (7). 
In our ansatz, we can explicitly obtain
\begin{eqnarray}
\varepsilon_1 \approx - \frac{3 \lambda^6}{16 \pi} 
\left [ \sin 2 (2\omega - \zeta) - 2x (1 + x^2) \sin \omega \right . ~
\nonumber \\
\left . + x^2 z \sin 2\omega \right ]
\left [ z \left ( 1 + x^2 + z^2 - 2xz \cos\omega \right ) \right ]^{-1}  .
\nonumber
\end{eqnarray}
For illustration, we adopt the specific parameter space
chosen above to evaluate the size of $\varepsilon_1$.
The result is $\varepsilon_1 \approx - 5.2 \times 10^{-6}$.
To translate this CP asymmetry into the lepton asymmetry $Y_{\rm L}$
and the baryon asymmetry $Y_{\rm B}$, 
it is necessary to calculate the dilution factor $d$ appearing in Eq. (4).
Note that $d$ depends closely on the following quantity:
$K_{\rm R} \equiv [U^{\rm T} M^\dagger_{\rm D} M_{\rm D} U^*]_{11}
M_{\rm Pl}/(8 \pi v^2 \cdot 1.66 \sqrt{g^{~}_*} M_1)$
with $M_{\rm Pl} \approx 1.22 \times 10^{19}$ GeV,
which characterizes the out-of-equilibrium decay rate of $N_1$.
Making use of the typical inputs taken above, we arrive at
$K_{\rm R} \approx 73$. The dilution factor $d$ can then be
calculated with the help of an approximate parametrization \cite{Kolb} 
obtained from integrating the Boltzmann equations
(for the range $10\leq K_{\rm R} \leq 10^6$):
$d \approx 0.3/[K_{\rm R} (\ln K_{\rm R})^{0.6}] \approx 1.7 \times 10^{-3}$.
Finally we get a very instructive prediction for the baryon asymmetry 
of the universe from Eq. (4):
$Y_{\rm B} \approx 4.7 \times 10^{-11}$. One can see that this result is 
consistent quite well with the observational value of $Y_{\rm B}$ 
quoted in Eq. (1).

One may go beyond the typical parameter space taken in this talk
to make a delicate analysis of all measurables or observables.
It is remarkable that we
can quantitatively interpret both the baryon asymmetry 
of the universe and the small mass-squared differences and large mixing 
factors of solar and atmospheric neutrino oscillations. 
In this sense, our ansatz is a {\it complete} phenomenological ansatz 
favored by current experimental and observational data, although 
it has not been incorporated into a convincing theoretical model.

This work was supported in part by the National Natural Science 
Foundation of China.

\end{document}